\documentclass[useAMS,usenatbib,onecolumn]{mn2e}
\usepackage{graphicx}
\usepackage{url}
\newcommand{\eqref}[1]{(\ref{#1})}

\newcommand{\cm}{cm$^{-1}$}

\title[ExoMol: molecular line lists ]{ExoMol: molecular line lists for exoplanet and other atmospheres}

\author[J. Tennyson and S. N. Yurchenko]{Jonathan Tennyson and Sergei N. Yurchenko \\
Department of Physics and Astronomy, University
College London, Gower Street, WC1E 6BT London, UK}

\date{Accepted XXXX. Received XXXX; in original form XXXX}

\pagerange{\pageref{firstpage}--\pageref{lastpage}} \pubyear{2012}

\begin{document}

\label{firstpage}

\maketitle

\begin{abstract}

The discovery of extrasolar planets is one of the major scientific
advances of the last two decades. Hundreds of planets have now been
detected and astronomers are beginning to characterise their
composition and physical characteristics. To do this requires a huge
quantity of spectroscopic data most of which is not available from
laboratory studies. The ExoMol project will offer
a comprehensive solution to this problem by providing spectroscopic
data on all the molecular transitions of importance in the atmospheres
of exoplanets. These data will be widely applicable to other problems
and will be used for studies on cool stars, brown dwarfs and
circumstellar environments. This paper lays out the scientific
foundations of this project and reviews previous work in this area.

A mixture of first principles and empirically-tuned quantum mechanical
methods will be used to compute comprehensive and very large
rotation-vibration and rotation-vibration-electronic (rovibronic) line
lists. Methodologies will be developed for treating larger
molecules such as methane and nitric acid. ExoMol will rely on these
developments and the use of state-of-the-art computing.

\end{abstract}


\maketitle

\section{Introduction}
\label{sec:intro}

Most information on  the Universe around us has been gained by astronomers
studying the spectral signatures of astronomical bodies.
Interpreting these spectra requires access to appropriate laboratory
spectroscopic data as does the construction of associated radiative
transport and atmospheric models. For hot bodies the quantities of
atomic and molecular data involved can be very substantial: beyond
that which is easily harvested using only laboratory experiments. This
problem led, for example, to the establishment of the Opacity Project
\citep{OP1,OP2,Seat05} some 35 years ago with the explicit aim of calculating
all the necessary radiative data involving atomic ions which could be of
importance for models of (hot) stars. This project was introduced
by papers laying the scientific \citep{S87} and computational \citep{bbb87}
background for the project.

Stars cooler than our own Sun have significant quantities of molecules
in their outer atmospheres. These molecules have spectra which are, in
general, much richer than those of atoms and atomic ions. They thus
both dominate the spectral signature of the cool stars and provide
their major opacity sources which, in turn, determine their
atmospheric structures. There are even cooler  objects which are neither stars
nor planets, called Brown Dwarfs. These objects are largely
characterised and classified according to the molecular features in
their atmospheres.  The last decade has also witnessed a rapid
escallation in the number of planets orbiting other stars (exoplanets)
that have been identified.  This number is still increasing rapidly. So
far spectroscopic studies of exoplanets are limited in the number,
their wavelength coverage and, in particular, their resolution.
However it is already apparant from those studies available, which are
so far largely confined to hot gas giant planets, that analysing their results
will place similar demands on molecular line lists
to the requirements of cool
stars and brown dwarfs.  Modelling and interpreting the spectra of
these objects requires data appropriate for temperatures up to about
3000 K.

Since the first detection of sodium in an exoplanet \citep{cdn02},
exoplanet spectroscopy has made rapid advances.
However, even with a rather limited set of
molecules detected in exoplanet atmospheres, there remain serious
problems with laboratory data. For example, methane was detected in
HD189733b by \citet{svt08}, who lacked the necessary data to determine
its quantity; even the presence of methane in other objects remains
controversial \citep{shn10,jt495}.

Planets and cool stars share some common fundamental characteristics:
they are faint, their radiation peaks in the infrared and their
atmosphere is dominated by strong molecular absorbers. Modelling
planetary and stellar atmospheres is difficult as their spectra are
extremely rich in structure and their opacity is dominated by
molecular absorbers, each with hundreds of thousands to many billions
of spectral lines which may be broadened by high-pressure and
temperature effects. Despite many attempts and some successes in the
synthesis of transition lists for molecular absorbers,
reliable opacities for many important species are still lacking.

\begin{figure}
\centering
\includegraphics[scale=0.4]{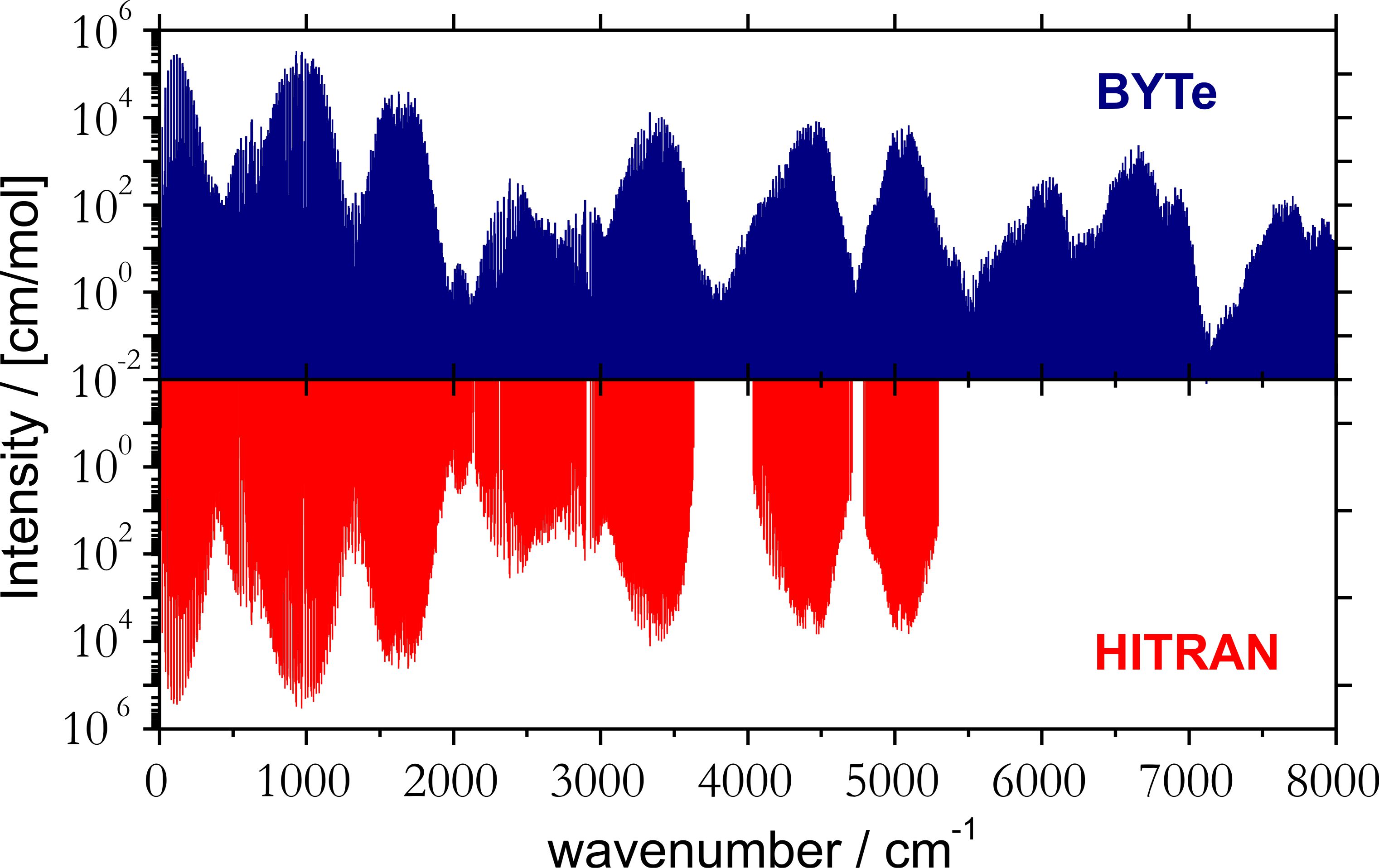}     
 \caption
{Room temperature ($T=296$~K) comparison of laboratory measured spectrum of
ammonia, as taken from the HITRAN database, with the line
list calculated using program ``TROVE'' by \citet{jt466}.}
\label{wholenh3}
\end{figure}

Determining line lists for hot molecules experimentally is difficult
because of (a) the sheer volume of data (maybe billions of lines), (b) the
difficulty in obtaining absolute line strengths in many cases, (c)
the need to have assigned spectra in order for the correct temperature
dependence to be reproduced, (d) the need for completeness, which requires a
large range of wavelengths; even at room temperature experimental line
lists are often far from complete, see Figure~\ref{wholenh3} for an example.
All this means that a purely empirical strategy is problematic.
Instead the plan is to build a reliable theoretical model for each
molecule of importance, based on a combination of the best possible
{\it ab initio} quantum mechanical treatment which is then validated by
and, in most cases, tuned using experimental data.

Molecular spectra, particularly for polyatomic species, rapidly become
extraordinarily rich at elevated temperatures, meaning that the data
requirement for a single triatomic molecule can outstrip the entire
Opacity Project dataset. Figure~2 illustrates the strong temperature-dependence
of the spectrum of water which requires many millions of line to simulate
at higher temperatures. Considerable effort has been expended in
constructing spectroscopic databases, such as HITRAN \citep{HITRAN} and
GEISA \citep{jt504}, which provide lists of molecular transitions
important at about 296~K. These are appropriate for modelling the
atmosphere of our planet and those of the other members of our solar
system. However the construction of accurate and complete databases
for higher temperatures has been much more partial with most high
accuracy studies concentrating on a single species. The present status
of this data is reviewed below.

This paper lays out the scientific foundations of a new project,
called ExoMol, which aims to systematically provide line lists for
molecules of key astronomical importance. These molecules have been
selected to be those most likely to be present in the atmospheres of
extra-solar planets. In practice they are of importance in many other
hot astronomical environments, particularly brown dwarfs and cool stars.
The ExoMol project aims to provide a comprehensive database for these
objects too. The following section summarises the presently available
line lists and illustrates the importance of these line lists
by considering some of problems they have been applied to. Section 3
considers the requirements for providing comprehensive data. The molecules
concerned are categorised on physical grounds and appropriate methodologies
are suggested for each class of problem. Section 4 gives conclusions
and perspectives.


\begin{figure}
\centering
\includegraphics[width=0.6\linewidth]{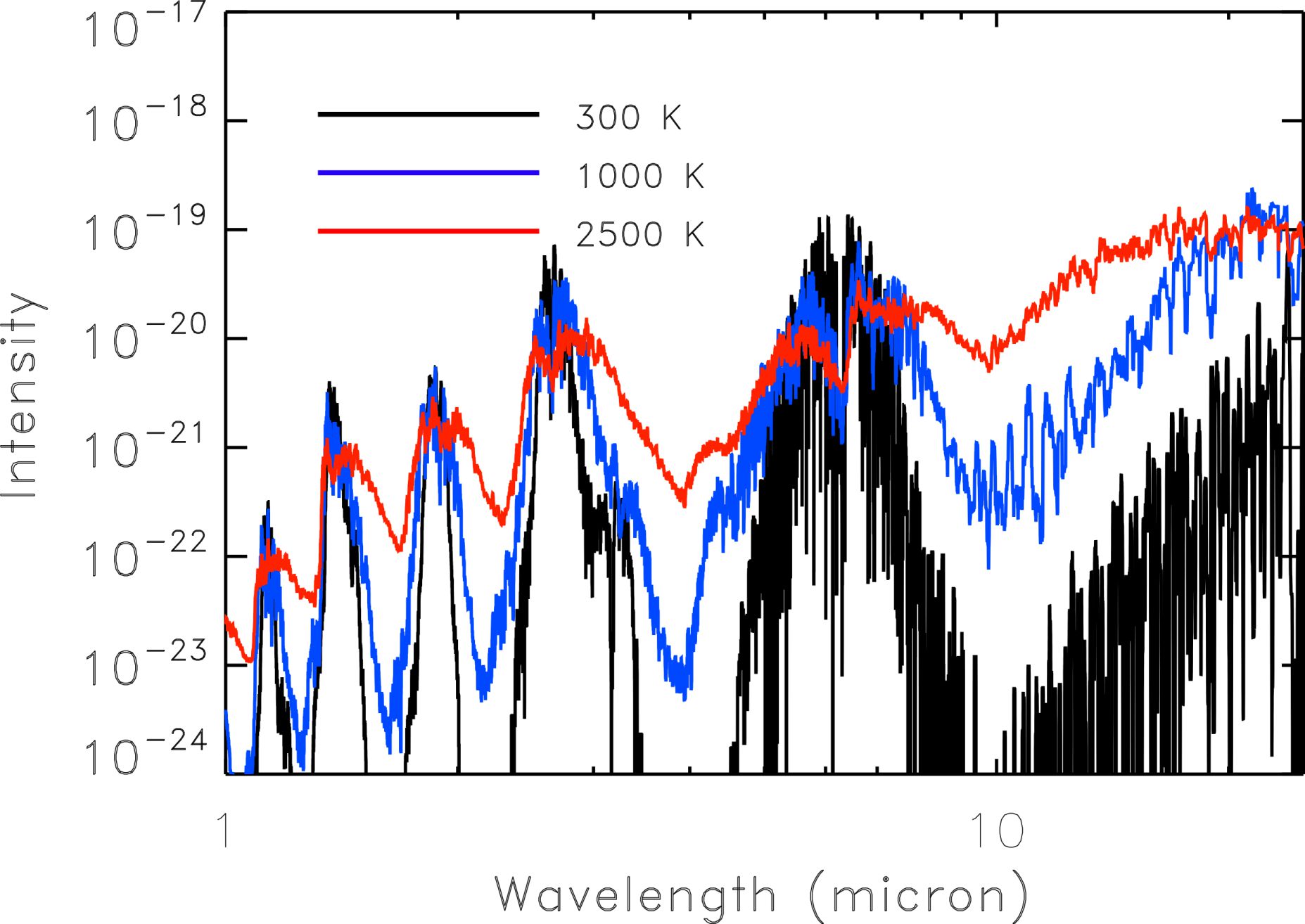}     
\caption
{Absorption spectra of H$_2$O given by BT2 \citep{j415} for $T$ = 300, 1000,
and 2500 K.}
 \label{f:water}
\end{figure}

\mbox{}\\

\section{The current situation}

Astronomers interested in molecular line lists use a number of
collected data sources. Besides HITRAN and GEISA mentioned above, the
JPL \citep{jpl} and CDMS \citep{cdms} databases provide comprehensive
molecular line lists for wavelengths longer than 30~$\mu$m. However these databases are
aimed at the cool interstellar medium rather than hot sources. HITEMP
in both its original \citep{hitemp95} and recently
updated \citep{jt480} editions carries data
appropriate for modelling molecular spectra at elevated temperatures,
but only for five species. Kurucz has extended his well-used atomic
opacity tables with
data for a number of molecules \citep{Kurucz} but the data for the
majority of molecules are approximate and the list of molecules far
from complete. Similarly there are partial lists of diatomic opacities
provided by the UGAMOP database at the University of Georgia (see
www.physast.uga.edu/ugamop/) and the RADEN databank at Moscow State University
\citep{raden}.
The SCAN database also contains line lists for a few diatomics and
triatomics \citep{Jor96}. In summary, while there are a number
of sources of molecular line list data, none of them can
be considered complete, especially for work at elevated temperatures.

In their review of brown dwarf and very low mass star atmospheres,
\citet{aha97} found that the majority of molecular opacities available
were based on statistical or similarly approximate treatments. Indeed
they quote no case where they considered the available molecular
line lists to be adequate. This situation has improved somewhat since
1997; Tables~\ref{diatomic} and \ref{polyatomic} summarise what we
believe to be the current situation for line lists of hot diatomic and
polyatomic species respectively.

\subsection{Diatomics}

\begin{table*}
 \centering
\caption{ Recommended available line lists for hot diatomic molecules.
Given is the main reference, method: experimental (expt), {\it ab initio} (ai)
or semi-empirical (semi), isotopologues other than the main one,
completeness up to a given estimated temperature ($T^{\rm max}$), number
of lines in the line list ($N$) and the electronic source of the data, if available.}
\small
\begin{tabular}[hl]{llclcrl}
\hline
Molecule  &  Ref  &  Method &  Isotopologues  &  $T^{\rm max}$$^{c}$ &  $N$ & Available$^d$ \\
\hline
C$_2$  &  \citet{Kurucz}                          &  semi  &     &    5200~K  &  3459595  &  Kurucz \\ 
CH     &  \citet{Kurucz}                          &  semi  &     &    3000~K  &  71591    &  Kurucz \\  
CN     &  \citet{Kurucz}                          &  semi  &     &    5000~K  &  1644597  &  Kurucz \\  
CO     &  \citet{jt480}                           &  expt  &     &    4000~K  &  113631   &  HITEMP \\
CaH    &  \citet{03WeStKi.CaH}                    &  semi  &     &    3500~K  &  89970    &  UGAMOP \\ 
CrH    &  \citet{02BuRaBe.CrH}                    &  semi  &  $^{50}$Cr,$^{53}$Cr,$^{54}$Cr &   1300~K & 13824  &  Bernath  \\ 
FeH    &  Bernath and co-workers$^a$              &  semi  &  $^{54}$Fe,$^{57}$Fe,$^{58}$Fe &   1600~K & 116300 &  Bernath  \\ 
HD$^+$ & \citet{jt506}                            &  ai    &     & all    & 10120     & ExoMol   \\
HeH$^+$& \citet{jt347}                            &  ai    & all$^e$ &  all   &    8573     & ExoMol \\
LiCl   & \citet{04WeKiHa.LiCl}                    &  semi  &     & 4000K      & 3357811   & UGAMOP \\  
LiH    & \citet{jt506}                            &  ai    &     & all        & 18981     & ExoMol\\
LiH$^+$& \citet{jt506}                            &  ai    &     & all        &  329      & ExoMol\\
MgH    &  Weck and co-workers$^b$   &  semi  &     &     1300~K &   23315   & UGAMOP \\ 
OH     &  \citet{jt480}                           &  expt  &     &    4000~K  &  41577    & HITEMP \\
NH     &  \citet{Kurucz}                          &  semi  &     &    3000~K  & 36163     &  Kurucz\\ 
NO     &  \citet{jt480}                           &  expt  &     &   4000~K   &  115610   &   HITEMP  \\
SiH    &  \citet{Kurucz}                          &  semi  &     &   3000~K   &  78286    & Kurucz\\ 
SiO    &  \citet{93LaBaxx.SiO}                    &  semi  &     &            &           &   \\
SiO    &  \citet{Kurucz}                          &  semi  &     & 5300~K     & 1827047   & Kurucz\\   
TiH    &  \citet{05BuDuBa.TiH}                    &  semi  &     &   1800~K   & 199073    &  Bernath  \\ 
TiO    &  \citet{98Scxxxx.TiO}                    &  semi  &     &  6200~K    & 37744499  & Kurucz \\ 
\hline
\label{diatomic}
\end{tabular}
\\
\flushleft
$^a$ \citet{03DuBaBu.FeH,10WeReSe.FeH,10HaHiBa.FeH}\\
$^b$ \citet{03WeScSt.MgHline,03WeScSt.MgHcontinuum,03SkWeSt.MgH,03WeStKi.MgH} \\
$^c$ $T^{\rm max}$ should be considered as a guide indicating the completeness of a line list in question, estimated from the 
maximal energy  $E^{\rm max}$ of the line list and the following condition on the Boltzmann factor:
$\exp\left({-{E^{\rm max}}/{kT^{\rm max}}}\right) = 5\times 10^{-7}$. 
The latter is an empirical threshold that corresponds to $T^{\rm max}$ and $E^{\rm max}$  of the  BT2 line list~\citep{jt378} (see Table~\ref{polyatomic}).  'all' indicates that all bound-bound  transitions within the ground electronic state are given.  \\
$^d$ Data sources:\\
Bernath: http://bernath.uwaterloo.ca/XY, where XY is the chemical formula of the molecule\\
CDSD databank: ftp://ftp.iao.ru/pub/CDSD-4000\\
ExoMol project: www.exomol.com\\
HITEMP: http://www.cfa.harvard.edu/hitran/HITEMP.html\\
Kurucz CDs, http://kurucz.harvard.edu/\\
UGAMOP project:      http://www.physast.uga.edu/ugamop/\\
$^e$ `all' means all possible stable isotopologues.

\end{table*}

Table~\ref{diatomic} lists diatomic line lists which are published,
available and fairly complete. Thus, for example, we have omitted the
HF line list used by \citet{08UtArLe.HF} as there is no source for
this data or, indeed, any details on how it was calculated. Similarly
the recent AlO line list of \citet{11LaBexx.AlO} contains accurate,
measured line frequencies but no transition intensities. In
addition, the MARCS
model atmosphere code \citep{MARCS} contains
unpublished molecular line opacities
for a number of diatomic species.

It is interesting to consider some of the diatomics that have been
treated.  Only in a minority of cases, specifically CO, OH and NO
which all form part of the HITEMP database \citep{jt480}, have the
line lists been constructed essentially on the basis of experimental
data, see for example \citet{94Goxxxxxx.CO} and \citet{08BeCo.OH}.
A more typical and demanding situation is given by TiO.

TiO is a major opacity source in cool, oxygen-rich stars
\citep{aha97}. It is an open shell system with several low-lying
electronic states which can absorb at near-infrared and red
wavelengths, that is close to the radiation peak in a cool star. A
number of theoretical studies provided at least partial line lists for
this system \citep{94Joxxxx.TiO,98Plez.TiO,98AlPexx.VO}.  At the same time there have been
several detailed experimental spectroscopic studies on the system
\citep{91GuAmVe.TiO,91SiHaxx.TiO,95AmAzLu.TiO,95KaMcHe.TiO,96RaBeWa.TiO}.
\citet{98Scxxxx.TiO} combined these studies and data from earlier
laboratory spectra
\citep{74Lixxxx.TiO,79HoGeMe.TiO,80GaBrDa.TiO,85BrGaxx.TiO} with state
of the art {\it ab initio} calculations to give a comprehensive TiO
line list containing 37 million lines.  This line list and the
corresponding TiO opacity was found to give a very good representation
of the TiO absorption in cool stars \citep{ahs00} and is now widely
used. Schwenke's TiO line list is the largest available for a diatomic
by more than an order-of-magnitude.

The next largest available diatomic line list is that for C$_2$ which
is one of a number of diatomics species for which line lists have
been provided by Kurucz. Recently, however, there have several new
experimental measurements on this system, including the characterisation
of entirely new, low-lying electronic bands
\citep{06KoReMo.C2,07JoNaRe.C2,07TaHiAm.C2,09NaJoPa.C2,10BoKnGe.C2,11BoSyKn.C2}. This work
has been accompanied by significantly improved {\it ab initio} electronic
structure calculations \citep{07KoBaSc.C2,07ScBaxx,09NaJoPa.C2,11SchmBa}. Given its importance,  C$_2$ is one
of the species we aim to provide an updated line list for.

The important CN radical is represented only by the \citet{Kurucz} data in Table~\ref{diatomic}, which  is somewhat approximate.
We should note the recent experimental efforts  \citet{10RaWaHi.CN} and by \citet{11RaBexx.CN} offering accurate but only
partial information. More work is
therefore needed on this species.

Before turning to larger molecules it is worth considering the
FeH and MgH molecules. Both these molecules have been the subject of
experimental studies which have provided partial line lists.

In the case of FeH \citet{03DuBaBu.FeH} present results on rovibronic
transitions within the $F~^4\Delta$ -- $X~^4\Delta$ electronic band. The
available line list is based on measured transitions which give the
spectroscopic constants of rotational levels belonging to the $v = 0, 1,
2$ vibrational levels of the FeH $X$ and $F$ states; these are then
extrapolated to $v = 3$ and 4, and for $J$ $(= N + S)$ values up to 50.5,
where $v$ is the vibrational, $J$ is the total angular momentum,
$N$ is the rotational, and $S$ is the spin quantum numbers.  The
line list for this band therefore consists of experimental and
extrapolated term values for the 25 vibrational bands with $v \le 4$.
The line list of \citet{03DuBaBu.FeH} was verified and corrected, by scaling
the Einstein $A$-coefficients, by \citet{10WeReSe.FeH} using the
high-resolution spectra of red dwarf star GJ~1002.
\citet{10HaHiBa.FeH} provide a
line list for the FeH  $E~^4\Pi$ -- $A~^4\Pi$ electronic system
near 1.58 $\mu$m which combined measured frequencies with {\it ab initio}
calculation of the linestrengths.

MgH, along with CrH, is of potential interest for measuring the
presence of deuterium in brown dwarfs \citep{08PaHaTe.CrH}, the
so-called deuterium test \citep{bzr99}. Extensive experimental
studies of MgH
electronic spectra have been performed by \citet{03WeScSt.MgHline},
\citet{03WeScSt.MgHcontinuum} and \citet{03SkWeSt.MgH}.
These have been used to compute the complete line list for the
$B^{\prime}$~$^2\Sigma^+$ -- $X~^2\Sigma^+$ system of $^{24}$MgH.
The list includes transition energies and oscillator strengths over
the  11,850 -- 32,130 \cm\ wavenumber range, for all possible allowed
transitions from the ground electronic state vibrational levels
$v^{\prime\prime}\le 11$. This list was computed using the best
available {\it ab initio} potential energies and dipole transition moment
function, with the former adjusted to account for experimental data.
The status of CrH is somewhat similar to this.

It is clear from the above that a mixed experimental and theoretical approach
has been the most successful so far. Other experimental datasets are
available, for example a very extensive study has recently been completed
on NiH \citep{09VaRiCr.NiH,12RoCrRi.NiH}, and these will provide an appropriate starting point for
further line lists.

\subsection{Polyatomic molecules}


\begin{table*}
 \centering
\caption{Recommended available line lists for hot polyatomic molecules. All line lists are theoretical and
designed to be complete up an estimated maximum temperature, $T^{\rm max}$;
the number of lines in millions for the main isotopologue only. For data sources see
footnote $d$ to Table~\ref{diatomic}.}
\small
\begin{tabular}[hl]{lllcrl}
\hline
Molecule &  Ref           &   Isotopologues  &    $T^{\rm max}$ &  $10^{-6}N$ & Available\\
\hline
H$_3^+$  &  \citet{jt181}       & H$_2$D$^+$ \citep{jt478}  &  3000~K  &   12  & ExoMol \\
H$_2$O   &  \citet{jt378}       & HDO \citep{jt469}         &  3000~K  &   503 & ExoMol \\
HCN/HNC  &  \citet{jt374}       &  H$^{13}$CN \citep{jt447} &  3000~K  &   240 & ExoMol \\
C$_3$    &  \citet{89JoAlSi.C3} &                           &  3100~K  &       &        \\
CO$_2$   &  \citet{11TaPe.CO2}  & all                       &  5000~K  &   626 &  CDSD  \\
NH$_3$   &  \citet{jt500}       &                           &   1500~K & 1014  & ExoMol          \\
\hline
\label{polyatomic}
\end{tabular}
\end{table*}

Table~\ref{polyatomic} summarises the available line lists for polyatomic
molecules. In the case were several line lists are available, only the
recommended one is given.

For polyatomic molecules the main methodology has been theoretical.
So far calculations have, of necessity, been performed piecemeal and
molecule-by-molecule. For key molecules many line lists may be
available; for example, at least seven lists are available  for hot water
\citep{jt143,wr92,jt197,ps97,JJS01,sp00,jt378}.  Studies have shown
significant differences between the use of different line lists (see
\citet{jt284} for example). It is clear that use of complete and
spectroscopically accurate line lists is important both for modelling
hot astronomical objects and for interpreting their spectra.

The recent CDSD-4000 CO$_2$ line list of \citet{11TaPe.CO2} extended their
earlier work, which was used in the 2010 edition of HITEMP \citep{jt480}, to higher temperatures.
This line list was constructed using effective Hamiltonians parameterised using
experimental data.
Carbon dioxide is a more rigid molecule that the other polyatomics considered in
Table~2 and thus a good candidate for treatment using effective Hamiltonians.
Very recent, high-temperature, emission experiments by \citet{11DePeRi.CO2} suggest
that the CDSD-4000 line list is indeed the best available at modelling high temperature
spectra but there remains some work to be done on this problem. We note that a new theoretical study using methods closer to those advocated here has
recently started \citep{12HuScTaLe.CO2}.

It is instructive to consider the breadth of applications of the calculated line lists,
many of which could not have been anticipated prior
to their construction.
The polyatomic line lists summarised in Table~\ref{polyatomic}
have all also been used to predict, analyse and assign
laboratory spectra of the species, especially at elevated
temperatures. These line lists also provide a source of cooling functions
\citep{jt489,jt506} and
high-temperature partition functions
\citep{jt169,jt263,jt304} which are important for a variety of astrophysics
problems.
In addition other applications can be summarised as follows.

\mbox{}\\
\noindent {\bf H$_3^+$}. \citet{jt181}'s line list  and  related partition function \citep{jt169} have been used:
\begin{itemize}
\setlength{\itemsep}{0.1mm}
\item to give all transition intensities for interpreting astronomical
observations since there are no laboratory absolute intensity
measurements for the spectrum of H$_3^+$;
\item to significantly improve models of cool white dwarfs stars \citep{brl97};
\item to resolve issues with the Jovian energy budget \citep{jt258};
\item to probe the role of H$_3^+$ in primordial cooling \citep{gs09};
\item to provide stability limits for giant extrasolar planets orbiting near their star \citep{kam07};
\item to model non-thermal rotational distributions of H$_3^+$ both in
the  interstellar medium \citep{oe04} and in storage ring experiments \citep{jt306,jt340}.
\end{itemize}

\mbox{}\\
\noindent {\bf Water}. The BT2 line list \citep{jt378} has been used:
\begin{itemize}
\setlength{\itemsep}{0.1mm}
\item to show an imbalance between nuclear spin and rotational temperatures
in cometary comae \citep{jt330,jt349} and assign a new set of, as yet unexplained, high energy
water emissions in comets \citep{jt452};
\item to detect and analyse water spectra in (a) Nova-like object V838 Mon \citep{jt357},
(b) atmospheres of brown dwarfs \citep{jt417} and (c) FU Orionis objects;
\item to calculate the refractive index of humid air in the infrared \citep{Mat07};
\item to detect water on transiting extrasolar planets, for which it
  was completely instrumental \citep{jt400} as other available line
  lists did not contain good enough coverage of the many weak lines
  that become significant absorbers at high temperatures to make this
  detection securely;
\item for high speed thermometry \citep{Kr07}, tomographic \citep{Ma09} imaging in gas engines and burners and input for models of jet engines \citep{Lindermeir20121575};
\item as input for an improved theory of line-broadening \citep{jt431};
\item to model water spectra in the deep atmosphere of Venus \citep{jeremy09};
\item to validate the data used in models of the earths atmosphere and in particular simulating the contribution of weak water transitions to the so-called
water continuum \citep{jt463}.
\end{itemize}


\noindent {\bf HCN/HNC}. \citet{85JaAlGu.HCN} showed that including HCN
opacity in their model of atmospheres of cool carbon-rich stars caused the  modelled atmosphere
to expand by a factor of 5 and lowered the gas pressure of
the surface layers by 1 or 2 orders of magnitude.
This finding did much to stimulate detailed work on molecular
line opacities. Subsequent line lists \citep{jt298,jt374} have treated the isomerising
HCN/HNC as a single species. They have been used
\begin{itemize}
\setlength{\itemsep}{0.1mm}
\item to detect HNC in the spectra of carbon stars \citep{jt321};
\item to constrain C and N abundances in of AGB stars \citep{mzv05};
\item for models of the thermochemistry of HCN \citep{jt304};
\item to assign a particularly extensive set of hot, laboratory HCN
\citep{11Mexxxx.HCN} and HNC \citep{11Mexxxx.HNC} spectra.
\end{itemize}

\mbox{}\\
\noindent
{\bf HeH$^+$}. This molecule had been neglected from standard models of helium-rich white
dwarfs (which rather surprisingly included both H$_2^+$ and
He$^+_2$, see for example \citet{stan94}). The line list of \citet{jt347} has been used
\begin{itemize}
\setlength{\itemsep}{0.1mm}
\item for models of the white dwarf stars, particularly helium-rich white dwarfs \citep{jt342};
\item to study the effects of early chemistry on the cosmic ray background \citep{sgp08};
\item as a starting point for calculations of end effects in the upcoming
KATRIN neutrino mass measurement experiment \citep{jt372}.
\end{itemize}

To add context to the methods discussed below, it should be noted that while
the H$_3^+$ and HeH$^+$ line lists are completely {\it ab initio}, those
for water, HCN/HNC, and NH$_3$ used laboratory measurements to improve
the procedure. For water and ammonia this came via the use of a
spectroscopically determined potential energy surface (PES), while the
HCN/HNC calculations replaced the calculated {\it ab initio} energy
levels with observed one where known.  Both of these procedures, plus
other methods discussed below, take advantage of laboratory high
resolution spectra.
In contrast, comparisons between dipole transitions
intensities computed using completely {\it ab initio} dipole moment surface
(DMS) and benchmark experimental studies have shown
that intensities calculated {\it ab initio} are competitive with, and
often more accurate than, laboratory intensity measurements even when they
are available \citep{jt413,jt509}.


\subsection{Scope of the ExoMol project}
A list of species that we plan to consider is given in
Table~\ref{species}.  This list is based on current demands for models
of exoplanets and brown dwarfs. However it is necessary to be flexible
since as the characterisation of exoplanets improves
additional molecular line lists are likely to be required.

Broadly speaking, it is possible
to separate the spectroscopic demands for hot Jupiters, which
have a reducing or hydrogen-rich chemistry, and super-Earth
type exoplanets, which can be expected to be oxidising or oxygen-rich.
In practice there will be other categories of exoplanets, such as warm
Neptunes; however the above division should be sufficient to identify
the species for which data are  required.

Although the atmospheres of other exoplanets are now starting
to be probed \citep{Bean2010,jt495},
those exoplanets that are currently the subject of spectroscopic
analysis are largely hot Jupiters.
For hot Jupiters there is already a major demand for a
high quality methane line list \citep{svt08}
and there are likely to be demands for
line lists of other hydrogenated species such as H$_2$S, PH$_3$,
acetylene, ethane and propane. For super-Earths, that is planets with
rocky cores, oxygen bearing species such as ozone and oxides of
nitrogen and sulphur are likely to be more important and also
will provide bio-signatures \citep{dhj02}.
Simulated remote spectra of habitable planets \citep{ksf10} suggest that
a variety of molecules including even nitric
acid could provide a possible bio-signature. Finally it is already
known from models of cool stars that open shell diatomics with
low-lying electronic states have clear atmospheric signatures and can
play an important role in determining the radiative transport
properties of the atmosphere. So far only for TiO \citep{98Scxxxx.TiO} is
there a satisfactory line list available.

Table~\ref{species} summarises the molecules that are thought likely
to be important in the atmospheres of extra-solar planets and cools
stars. They are classified according to anticipated atmospheric
chemistry, since this is radically different in systems which have no
heavy elements (such as bodies formed in the very early universe), are
oxygen-rich or are carbon-rich. These species are separated between
those for which satisfactory line list, extending to high temperature,
are currently available, and those for which line lists are needed. It
is clear that, except for primordial chemistries, the to-do list is
much the longer. Table~\ref{species} has been constructed from
the literature (eg \citet{sb07,fml08}) and as a result
of extended discussions with several scientists  involved directly or
indirectly in characterising exoplanets.

\begin{table}
\caption{Exomol list of molecules: Molecular line lists available, fuller lists are given
in Tables~\ref{diatomic} and \ref{polyatomic}, and planned
categorised by chemistry.}
\begin{tabular}{llll}
\hline
& Primordial & Terrestrial Planets & Giant-Planets \&\ Cool Stars\\
& (Metal-poor) & (Oxidising) &  (Reducing atmospheres)\\
\hline
Already  & H$_2$, HD$^+$,  LiH, LIH$^+$ & OH,  CO$_2$,  O$_3$, NO & H$_2$,  CN, CH, CO, CO$_2$, TiO, YO,\\
available&     HeH$^+$, H$_3^+$, H$_2$D$^+$     & H$_2$O, NH$_3$& HCN/HNC, H$_2$O, NH$_3$ \\
\hline
ExoMol   &  BeH & O$_2$, CH$_4$, SO$_2$  & CH$_4$, PH$_3$, C$_2$,  C$_3$, HCCH, \\
         &      & HOOH, H$_2$CO, HNO$_3$ & C$_2$H$_6$, C$_3$H$_8$, VO, O$_2$, AlO, MgO, \\
         &      &                        & BeH, CrH, MgH, FeH, CaH, AlH, SiH, NiH, TiH\\
\hline
\end{tabular}\label{species}
\end{table}

There are significant
differences in the physical processes that need to be modelled for different molecules. These
are addressed in the next section. Based on the underlying physics
that needs to be considered, the problems
can classified as (1) diatomics, (2) triatomics, (3)
tetratomics, (4) methane and (5) larger molecules. Special techniques will
be required in each case. A final topic will focus on the content, construction
and use of the ExoMol database itself.

%

\section{Methodology}

Most of the molecules listed in Table~\ref{species} are chemically
stable and only undergo electronic transitions in the ultraviolet. For
these systems it is necessary to consider in detail pure rotational
and vibration-rotation transitions within the ground electronic
state. However the list also contains a number of open shell diatomic
species such as C$_2$ and FeH.
These molecules undergo electronic
transitions at near infrared or visible wavelengths; for these systems
it is necessary to consider electronic transitions
and hence excited electronic states.

\begin{figure}
\centering
\includegraphics[scale=0.5]{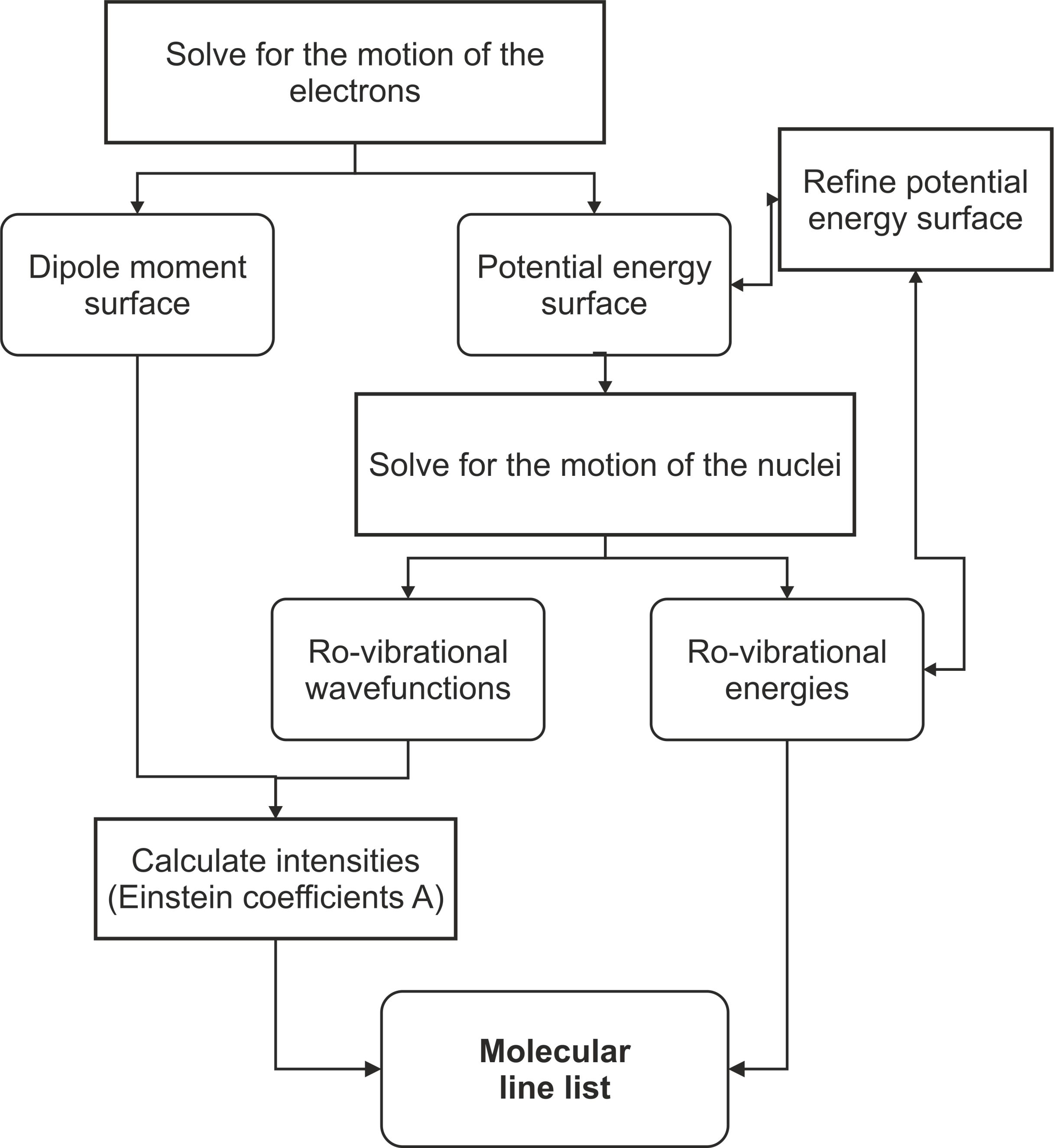}     
\caption
 {Schematic of the general method that will be used to produce
molecular line lists.}
 \label{method}
\end{figure}

Within the Born-Oppenheimer approximation, the calculation of line
lists of rotation-vibration transitions for a stable polyatomic
molecule can essentially be broken down into the following steps:
ground-state electronic structure calculations to give energies and
electric dipoles at a series of geometries; interpolation between
geometries to create PES and DMS;
nuclear motion calculations to provide energy levels and wave
functions; calculation of transition dipoles using the wave functions
and DMS. Predicted frequencies which arise from such
a purely {\it ab initio} procedure are only accurate enough for
present purposes for electronically very simple systems. It is
therefore necessary to improve these frequencies using experimental
data. This can be done in one of three ways:\\
a. {\it A priori} by tuning the PES by comparison with
the results of laboratory high resolution spectroscopic studies. We
have developed new procedures for this \citep{jt423,jt503} which are
both efficient and retain the
predictive nature of the underlying {\it ab initio} PES.\\
b. {\it Post hoc} by replacing calculated energy levels with observed
ones. As energy levels are not observed directly we will rely on the
MARVEL inversion procedure \citep{jt412}
which has been successfully used for water isotopologues \citep{jt454,jt482}.\\
c. During the calculation: since most of the error is in the
vibrational not the rotational energies \citep{jt205}, using empirical
vibrational band origins can significantly improve predicted
frequencies for all transitions in the band. The nuclear motion
program TROVE \citep{trove-paper} contains the facility to replace
predicted band origins with empirical ones part way through the calculation.
This facility was used in constructing the recent BYTe line list for NH$_3$
\citep{jt500}.
In practice all three methods
will be used, often in combination. Figure~\ref{method} summarises
our general methodology.

For each line list the following components are
required:\\
(i) nuclear motion model, implemented in a computer program,
for accurate calculations of rotation-vibration energies and wave functions,\\
(ii) accurate PES and DMS,\\
(iii) a
computational procedure for intensity simulations based on the results
of the nuclear motion calculations.

It is generally accepted that {\it ab initio}
DMS computed at high levels of theory provide very reasonable
description of the intensities, better for example, than can be obtained
by attempting to fit a DMS to measured intensities \citep{jt156}.
Moreover, such properly obtained {\it ab initio} intensities are, in
all but a few cases, superior to data provided by experiment
\citep{jt509}.

{\it Ab initio}  PES, however, cannot
deliver rotation-vibration energies with sufficiently high accuracy.
It is therefore common to empirically refine {\it ab initio} PESs by
least-squares fitting to experimental energies or frequencies to give
a ``spectroscopic'' PES; such potentials can provide theoretical
line positions with near-experimental accuracy. When performing such
fits it is important to prevent the refined surface from distorting
into unrealistic shapes in regions not well characterised by the
experimental data. To this end we impose an additional constraint
requiring that the refined PES remains relatively close to the
underlying {\it ab initio} PES \citep{jt423}. Technically this is done
by simultaneous fitting of the potential parameters function both to
the experimental (rotation-)vibration energies and to the
(lower-weighted) {\it ab initio} energies \citep{Yurch2003}; this is an efficient and, as yet,
not widely used procedure.

As mentioned above,
the  MARVEL
(Measured Active Rotation-Vibration Experimental Levels) procedure \citep{jt412})
provides a rigorous protocol
for extracting experimental energy levels from the observed data. So
far this protocol has largely been applied to water and the results
are only just being incorporated into line lists \citep{jt522}. We will make
extensive use of MARVEL during the ExoMol project; note that the ``Active''
means that the results can be updated as more or improved laboratory
measurements become available by simply including these measurements
and re-running the process. The original HCN/HNC line list \citep{jt298}
was already updated in this fashion \citep{jt374} and extended to
H$^{13}$CN/HN$^{13}$C \citep{jt447} using empirical energy levels,
although not ones obtained using MARVEL. This procedure took advantage of our
preferred data structure which seperates the line lists into energy and transitions
files, and which is discussed in Section~3.6.

There are essentially two different approaches to the construction of
the rotation-vibration Hamiltonian. Hamiltonians based on use of an exact kinetic energy (EKE) operator for the nuclear motion of the
molecule in question expressed in terms of geometrically defined
internal coordinates \citep{jt14}. EKE calculations are potentially very accurate, since the EKE approach can provide an exact (within the Born
Oppenheimer approximation) evaluation of the corresponding matrix elements \citep{jt511}. However, in order to achieve
this high accuracy, very substantial computer resources may be needed. Consequently, the EKE approach is the method of choice for triatomics
but its use has to be reviewed on a case-by-case basis for larger systems to ensure that the calculations are tractable.

As an alternative, the kinetic energy operator can be defined as a
Taylor series in some vibrational coordinates with expansion
parameters obtained numerically \citep{trove-paper}.  Such a non-EKE
approach can be adopted for the larger molecules, some tetratomics and
methane.  The rotation-vibration coordinates are chosen to minimise
the coupling between rotation and vibration. The vibrational
coordinates are taken essentially as the Cartesian displacements of
the nuclei from their positions when the molecule is at equilibrium.
For molecules with large amplitude motions, such as HOOH, we can follow
the HBJ method of \citet{hbj70} and expand the nuclear kinetic energy
operator in terms of the small-amplitude vibrational coordinates from
a flexible reference configuration, i.e. around the large-amplitude
coordinate. In our extensions of the HBJ theory, the eigenvalues and
eigenfunctions of the expanded Hamiltonians are determined
variationally by numerical diagonalisation of a matrix representation
of the expanded Hamiltonian. This procedure is implemented in the
program TROVE of \citet{trove-paper}.

For the largest molecules to be considered, those which have five or
more atoms (apart from methane), nuclear motion calculations will be
performed using new procedures. These will be based on the use of
more approximate methods, such as MULTIMODE \citep{multimode}, DEWE
\citep{Matyus-DEWE-2007} or an adaption of TROVE, which use normal coordinates and which are more
appropriate for fairly rigid molecules.  These calculations will not
be as accurate as those proposed for smaller systems and therefore
extensive tuning to experimental data will be required. Furthermore
the potentially huge number of lines required to reproduce a
line-by-line spectrum for these species at even moderately high
temperature is likely to be prohibitive.

For diatomic and triatomic systems we will generate
full line lists in which each rotation-vibration transition is explicitly calculated. For tetratomics
and methane, this will be also done where computationally feasible. Otherwise
vibrational band intensities and H\"onl-London factors will be used to give
transition intensities, while the calculated energy levels and hence frequencies
will be obtained from calculated, vibrationally averaged rotational
constants.

A number of steps will be undertaken to ensure  the accuracy of
our final line lists. For each system we will initially compute a less
comprehensive, low-temperature line list which can be checked for
reliability against available laboratory spectra. This step has proved
to be fundamental to the success of our recent line list
calculations. We note that in the case of methane there are extensive
laboratory data, particularly in the near infrared, which have so far
defied analysis. It is to be anticipated that our line lists will help
resolve these problems as  has been the case previously for
H$_3^+$, water and the HCN/HNC system.

\subsection{Diatomics}

For many diatomics it is necessary to consider electronic transitions.
Best results will rely on the availability of laboratory frequency
measurements.  Such measurements are, of course, more reliable than we
can calculate but are rarely complete, especially for elevated
temperatures.  Laboratory data are available for the majority of
systems such as the extensive dataset of NiH transition frequencies
and associated energy levels that have recently become available
\citep{12RoCrRi.NiH}. However, it is extremely difficult to construct
a complete, high temperature line list only from directly measured
data; for example the cited NiH spectra contain no usable information
on transition intensities. This makes the construction of a reliable
theoretical model essential.

Treatment of the nuclear motion problem for diatomics case is
relatively straightforward.  For uncoupled electronic states, we will
simply use the program LEVEL by \citet{lr07}. In more
complicated cases, where strong coupling between different electronic
states is important, it will be necessary to consider
this coupling explicitly in the calculation. \citet{95Marian.NiH,01Marion}
has developed a practical theory for such calculations.
A program to include couplings between electronic states
already exists \citep{ztf09} but will need to be extensively generalised
to cover the many different types of couplings that will be encountered
during the project. Given this, the main issue determining
the accuracy of the calculations is therefore one of
obtaining reliable potential energy curves, curve couplings and transition dipoles.

Our strategy here will be to start from high grade {\it ab initio}
methods: multi-reference configuration interaction (MRCI) expansions
based on full-valence complete active space, self-consistent-field
reference states utilising large Gaussian basis sets such as aug-cc-pV6Z to
resolve valence electron correlation effects. Core and core-valence
correlations and scalar relativistic energy corrections will also be
added. Such features are all standard in quantum chemistry programs
such as MOLPRO \citep{molpro}, Molcas \citep{molcas} and Columbus
\citep{columbus}.

High quality {\it ab initio} potential
energy and interstate coupling curves will provide the
starting point for further refinement. This will take two forms:
\begin{enumerate}
\item {\it Ab initio} methods will be used
to determine relativistic effects, and in particular relativistic
spin-orbit and other  couplings
between nearby curves.
\item {\it Spectroscopic} data will be used initially to test curves and then
to refine them to give curves that reproduce observed spectra. Much of this
can be done with program
DPotFit  \citep{dpotfit}, although this procedure
may need extending to deal with molecules whose electronic states are strongly
coupled.
\end{enumerate}

Work on diatomic line lists is actively underway and the following
paper \citep{jt529} reports line lists for the  $X~^2\Sigma^+$ states of
BeH, MgH and CaH.

\subsection{Triatomics}

As there are good line lists for the key triatomic species H$_2$O, CO$_2$,
HCN/HNC and H$_3^+$, relatively few triatomic line lists are planned.
Species to be considered include H$_2$S, C$_3$ and SO$_2$. The nuclear
motion problem for these species will be solved using the EKE DVR3D
triatomic code \citep{jt338}, which has already been extensively
adapted for the requirement of generating large line lists. For example
the algorithm to compute dipole transition intensities was both
reworked and parallelised to cope with the requirements of these
calculations.

A very accurate spectroscopic H$_2$S PES was constructed by
\citet{01TyTaSc.H2S}; this will provide a good starting point although
further work will be required to ensure that it remains reliable for
higher-lying states. Some work on the role of minor corrections to the PES
has already been performed \citep{jt295}. The
DMS for H$_2$S is less straightforward as H$_2$S has a known feature
that the dipole associated with the asymmetric stretch passes through
zero close to the equilibrium geometry making the DMS very sensitive
to the level of theoretical treatment used to model it. A DMS
calculated by \citet{02CoRoTy.H2S} purports to deal with this problem,
but in our tests has not been found to be uniformly reliable.
Therefore a new higher level theoretical treatment will be needed to
give a satisfactory solution to this problem.  The
procedures developed to produce an essentially exact DMS for water
\citep{jt424,jt509} will be used to determine the level of treatment
appropriate for H$_2$S. In particular, the DMS calculations will use finite
differences rather than expectation values which will allow us to test
the appropriate level at which to introduce relativistic and other ``minor''
effects prior to launching a full determination.

A C$_3$ line list was produced by \citet{89JoAlSi.C3}, one of
the very early ones to be produced. However this line list is no longer
accurate by modern standards.  C$_3$ is a complicated quasi-linear
system with an exceptionally flat PES which supports many
low-frequency bending modes \citep{97SpMeJe.C3}. These have been probed
via electronically excited states using stimulated emission pumping
\citep{89RoGoxx.C3,90NoSexx.C3,90RoGoxx.C3} or laser induced
fluorescence \citep{89Roxxxx.C3,93BaBrHa.C3}. Some preliminary work on
constructing a new C$_3$ line list has been performed \citep{jt407}
which was based on measurements \citep{06SaWexx.C3} and {\it ab
  initio} calculations \citep{04AhBaWe.C3} performed in Bristol.  A
first task will be to improve our current PES by performing a new {\it
  ab initio} calculation and then tuning to the available data.

The third triatomic for which a line list will be computed is
SO$_2$. This is a known constituent of solar system planetary
atmospheres \citep{nes90}.

\subsection{Tetratomics}

Two different codes will be used for treating the tetratomic nuclear
motion problem. TROVE \citep{trove-paper} has already been used to
compute an ammonia line list \citep{jt500} and will be used for
phosphine (PH$_3$) and formaldehyde (H$_2$CO). However this code is
not appropriate for systems which probe linear geometries such as the
linear acetylene (HCCH). For
this molecule the EKE code WAVR4 \citep{jt339} will be employed;
indeed acetylene was one of the molecules the code was originally
developed to treat \citep{jt346}.

A preliminary acetylene line list was computed by \citet{jt479}.  This
calculation gave reasonable results for spectra simulated using
vibrational (i.e. $J=0$) wave functions and vibrational band intensities
\citep{jt121} with the rotational fine structure given by
vibrational-state dependent rotational constants, and intensities
computed using H\"onl-London factors.  We will aim to produce an
improved line list based on a fully-coupled rotation-vibration calculation
but first further work will be required on both the PES and DMS. In
this context we note the recent emission spectra of hot acetylene
obtained by \citet{11MoGeBe.HCCH}.
In addition we will produce a line list for HOOH which is naturally treated
using the diatom-diatom coordinate option available in WAVR4.

Phosphine should be amenable to the the same treatment as ammonia and
has already been the subject of preliminary, low-temperature studies
\citep{06YuCaTh.PH3,08OsThYu.PH3,jtxxx}. Formaldehyde is the final
tetratomic planned. This has  lower symmetry than the other
tetratomics discussed so will be computationally the most demanding.
However a high quality, spectroscopically-determined PES has recently
been developed for this molecule \citep{11YaYuJe.H2CO} which will make
and excellent starting point for a line list calculation.

\subsection{Methane}

The detection of methane in exoplanet HD189733b \citep{svt08}
was notable for the failure to determine the actual quantity due to
the lack of appropriate laboratory data. Similar problems dogged
studies of the impact of comet Shoemaker-Levy 9 with Jupiter
\citep{jt198}, and also the interpretation of spectra of brown dwarfs
\citep{02HoHaAl.CH4}, where the desperate resort of modelling them
using methane spectra taken from solar system gas giants and Titan
has been used \citep{96GeKuWo.CH4}. Methane is,
of course, also an important greenhouse gas as well as being a
constituent of many flames. There has been for some time, a clear and
pressing need for a comprehensive database of methane transitions.
However this is a seriously challenging problem involving the
calculation of many billions of vibration-rotation transitions.
Advances in computer power and nuclear motion treatments mean that it
is becoming technically possible to contemplate a full and systematic
computational solution to the methane opacity problem.

There has already been some work in this direction.
\citet{02Scxxxx.CH4} performed some preliminary studies with a view
to computing a line list. More recently
\citet{09WaScSh.CH4} did compute a line list using the code MULTIMODE
\citep{multimode}. However this line list has neither the
number of transitions nor  the accuracy required for models of
exoplanet spectra, or indeed the other applications anticipated here.
Currently the main source of methane spectra are HITRAN \citep{HITRAN},
which is only really appropriate for temperatures below 300~K and is then
still not complete, or the low-resolution PNNL database \citep{sjs04}.
There has recently been significant and coordinated experimental
activity to try to understand methane spectra and create corresponding line lists
\citep{09AlBaBo.CH4,11WaMoKa.CH4}, in particular to aid the interpretation
of Titan spectra.

Methane is a 10 electron system like water (and ammonia): for water
there are well developed procedures for obtaining an ultra-high
accuracy PES \citep{jt309}; application of these
procedures should be easier for methane since it is possible to use
the faster coupled-clusters approaches, such as CCSD(T), instead of
MRCI because of the simpler topology of its PES.
Methane has nine degrees of vibrational freedom compared to water
which only has three and will therefore require the calculation of
significantly more geometries.  However methane's symmetry reduces
this number by a factor of 24 and use modern computers should allow us
to calculate upwards of 50000 points in a few months using MOLPRO
\citep{molpro} even with no frozen core, a large (6Z level) or F12
basis set \citep{hmp10} while also including allowance for relativistic
and adiabatic corrections. This potential can be improved using the
extensive experimental datasets referenced above.

The much more difficult steps are the calculation of the
12-dimensional rotation-vibration wave functions and the subsequent
calculation of all the associated transition intensities: it is to be
anticipated that the final line list will comprise many billions of
transitions. There are two possible strategies for solving this
problem. In both cases the vibrational calculations will be performed
with an upgraded and parallelised version of TROVE
\citep{trove-paper}. A strong point of TROVE is the automatic and
general treatment of symmetries; this is important not only because
maximising the use of symmetry will help to keep the calculation
tractable but also because symmetry is necessary to get nuclear spin
effects correct when generating spectra.

The comprehensive solution is to use  TROVE
to simply compute all possible
vibration-rotation transitions directly. The calculation of highly
rotationally excited states ($J$ up to 40) and, even more so, the
calculation of huge lists of dipole transitions are computationally
demanding in the extreme. It is unclear yet, both because the necessary
benchmark calculations need to be performed and also because
of uncertainty about what computer
power will be available,  whether this approach will be completely
feasible or only  so in part.

A more pragmatic, but still reliable, approach is to follow that
already used for the preliminary acetylene calculations \citep{jt479}.
Well-converged wave functions from a $J=0$
(i.e. vibration only) calculation will be used to compute (a)
vibrational band intensities and (b) vibrational-state dependent
rotational constants. The rotational constants will be used to
generate the required energy levels and transition
frequencies. Vibrational band origins will be combined with so-called
H\"onl-London factors (well known for the high-symmetry methane
molecule) to give the intensity of individual rotation-vibration
transitions.

\subsection{Larger molecules}

A characteristic of hot molecules is that their spectra become very
congested with many blended lines. For heavier molecules this limit
can be reached at room temperature. Thus, for example, the HITRAN
database \citep{HITRAN}
stores  all data on species with four or more heavy atoms
(about 30 species) as cross-sections rather than fully resolved line
lists. However cross-section data are only applicable at the temperature
of the measurement, usually room temperature. This is a severe
disadvantage which makes the data inflexible. For example, it would be
very difficult to use this cross-section data in atmospheric models of
an earth-like planet even when there is only a relatively small
temperature differential to earth, say for a super-earth at 350~K.

Conversely the use of variational procedures such as the one outlined
above for these heavy systems would be both computationally very expensive
and of much lower accuracy than would be required. We therefore propose
a  rather more empirical approach to address this problem.

The \citet{Wat68} Hamiltonian provides a general formulation for
the nuclear motion of polyatomic systems which is particularly
well-suited for use with semi-rigid molecule. This Hamiltonian
uses normal modes and, for well behaved systems it is possible to
simplify calculations by neglecting high-order coupling terms.
This  is the basis of the MULTIMODE approach of \citet{multimode}
and DEWE
\citep{Matyus-DEWE-2007};
TROVE can also be adapted to work in this fashion.
MULTIMODE has proved capable of giving reasonable
results for the rotation-vibration spectra of relatively large
molecules \citet{wbb08}.

Calculations based on Watson's Hamiltonian will be
 used to model room temperature spectra of
the species of interest, namely HNO$_3$, C$_2$H$_4$, C$_2$H$_6$ and
C$_3$H$_8$. The calculations will use {\it ab initio} potentials
largely represented by expansions about equilibrium in terms of force
constants and higher-order terms. These constants will be calculated
{\it ab initio} as derivatives at equilibrium, empirically-determined
if available or a mixture of the two. For some large amplitude modes,
such the CH$_3$ rotations, full potentials will be used. The PES
surface and, if necessary, the associated dipole moments will be
systematically tuned to reproduce the measured room temperature cross
sections for each system.  Initially this will have to be done
essentially by trial and error. However with several systems to work on it
is anticipated that we should be able to develop systematic procedures
for this tuning.

Having developed satisfactory room temperature models for each system,
calculations will be repeated at a grid of temperatures to give
temperature-dependent cross-sections. Experimental data such as
that by \citet{10LoBoLo.C2H4} for C$_2$H$_4$ will be used to either confirm the model or to
provide further input for an improved tuning procedure.

The final output of this model will be cross-sections, since at higher
temperatures the number of individual lines will simply be vast and
there is little prospect of high resolution spectra of these species
being fully resolved in astronomical observations in the near
future.

\subsection{Partition functions}

Partition functions are important for models of hot molecules and not
altogether straightforward to compute. Extensive
compilations of partition functions of astrophysically important
species have been made by \citet{81Irxxxx.partfunc} and
\citet{84SaTa.partfunc}. These compilations are comprehensive but do
not cover all the molecules to considered in the ExoMol project; the
partition function values themselves could also, undoubtedly, be improved
at higher temperatures. 
Partition functions which can be considered reliable over
an extended temperature range have been constructed for a number
of polyatomic molecules including H$_3^+$
\citep{jt169}, water \citep{jt263}, HCN/HNC \citep{jt304} and
recently acetylene \citep{11AmFaHe.HCCH}.

In general the partition function is given by
\begin{equation}\label{e:partfunc}
    Z(T) = \sum_{i}  g_i e^{ - c_2 E_i/T },
\end{equation}
where $c_2$ is the so-called second radiation constant and is appropriate when the energy, $E_i$, is given as a term value in cm$^{-1}$,   $T$ is the temperature
and $g_i$ is the statistical weight factor. The statistical weight deserves a special comment.

If the hyperfine structure to be unresolved the statistical weight factor is given by
\begin{equation}\label{e:stat-weight:N}
    g_i = (2S+1) \; g_{\rm ns}^{(i)} \; (2 N_i+1),
\end{equation}
where $S$ is the total electronic spin angular momentum, $g_{\rm ns}^{(i)}$ is the state 
dependent nuclear statistical weight and $N_i$ is the rotational angular momentum of the nuclei of the $i$th state. 
If the spin dependent states are resolved the statistical weight factor becomes 
\begin{equation}\label{e:stat-weight:J}
    g_i = g_{\rm ns}^{(i)} \; (2 J_i+1),
\end{equation}
where $J_i$ is total angular momentum of the $i$th state ($\bf{J}=\bf{N}+\bf{S}$). In Eqs.~(\ref{e:stat-weight:N}) and (\ref{e:stat-weight:J}) 
the hyperfine structure is assumed to be unresolved. 
We follow the HITRAN convention \citep{HITRAN-A} and include the entire
nuclear statistical weight $g_{\rm ns}^{(i)}$ of the molecule
explicitly in $g_i$ and hence the partition function $Z(T)$. For
example, for BeH where $^9$Be and H have nuclear spins $3/2$ and
$1/2$, respectively, $g_{\rm ns}^{(i)} = 8$.  If, as can be assumed for BeH
\citep{jt529}, the spin-rotation coupling for the ground electronic
state $X~^2{}\Sigma^+$ is unresolved, then each rovibronic states $i$ is assumed to be doubly ($2S+1$) degenerate. 
According to Eq.~\eqref{e:stat-weight:N} the statistical weight factor of BeH is then
given by $g_i = 2 \times 8 \times (2 N_i+1) = 16\;(2 N_i+1)$. If the spin dependent states are resolved 
then according to Eq.~\eqref{e:stat-weight:J} $g_i = 8\;(2 J_i+1)$. For the $^{12}$C$^{12}$C molecule whose nuclear spin is 0, the statistical weights $g_{\rm ns}$ of the symmetric $s$ and antisymmetric $a$ rotational levels are 1 and 0, respectively. In the case of the $X^1{}\Sigma^+$
ground electronic state of this molecule $g_{\rm e} = 1$ and the statistical weight factors $g_i$ are $(2 J_i+1)$ and 0 for the $s$ and $a$ states, respectively.

We note that the partition function, $Z$ can be used estimate the
completeness of a line list as a function of temperature. This can be
done by comparing the ratio of the partition function computed by
summing over all lower-state energy levels used to compute the line
list to the accurate partition function \citep{jt181,jt378}. Indeed
this ratio can even be used make corrections for the missing
contribution to, for example, the cooling function \citep{jt181}.

\subsection{The ExoMol database}

The backbone of the ExoMol database will be line lists of transitions. However
the database will include a variety of associated and ancillary data. These
will include energy levels, partition functions, cooling functions and
cross sections.

The large amount of data produced in the above calculations, both
completed and anticipated, requires the development of strategies for
data handling and distribution. Small line lists, which is likely include most
diatomics, a simple line list will be stored. However for larger line lists we
will employ a data structure which involves
organising our final line list into two files: an energy file and a
transitions file. The energy file will contain energy levels for each
state combined with a number for its position in the file and quantum
number assignments, both rigorous and approximate. The transitions
file will be arranged in ascending frequencies and will list only the
number of the upper and lower state for each transition plus the
associated Einstein-A coefficient. This provides a very compact means
of representing the data which is essential for efficient use of
storage. However, it is likely that the transition file will still
need to be split into frequency bins for ease of distribution.

This data structure has other important advantages.  It will also
allow us to actively update the energy file with measured
rotation-vibration energy, or indeed improved approximate quantum
numbers, ensuring the best possible  for each
transition.  Indeed \citet{jt447} turned an
H$^{12}$CN/HN$^{12}$C line list into one for H$^{13}$CN/HN$^{13}$C by
replacing the energy file with one appropriate for the $^{13}$C; this
approach relied on the not unreasonable assumption that the Einstein-A
coefficients do not change significantly between the two
isotopologues.

The sheer volume of data contained in the line lists makes them fairly
tricky to use. In practice most codes will use some sort of opacity or
importance sampling technique to identify key transitions and to
discard the rest. We have, however, constructed a set of
zero-pressure, temperature dependent cross sections for the key
species studied so far \citep{jt537}. The purpose of these is not to
replace the underlying line lists, whose use will remain necessary for
detailed studies and analysis of high resolution spectra, but to allow
the effects of adding a species to a model to be quickly and
efficiently tested.  The use of these cross sections
avoid the issues of handling huge line lists at the price of assuming
local thermodynamic equilibrium (LTE) and some loss of flexibility.
 The issue turning the line lists into cross sections will
be discussed elsewhere \citep{jt537}.

The results of the calculations outlined above will be a comprehensive
database of molecular transitions. The ExoMol database, see
\url{www.exomol.com}, will include not only the line lists, cross
sections cooling functions, partition functions and other data
generated during the project, but will also provide access those
already available. The database is web-based and our aim will be to
integrate it into the Virtual Atomic and Molecular Data Centre (VAMDC)
project \citep{VAMDC}. VAMDC data storage is based on the use of
XSAMS. XSAMS (XML Schema for Atoms, Molecules and Solids)
\citep{xsams} is an XML based data storage protocol which has been
designed by the International Virtual Observatory Alliance (IVOA) to
meet the needs of astronomers who wish to describe or access molecular
(and other) data in distributed datasets world-wide.

XSAMS is both flexible and intuitive, making data manipulation and
interpretation significantly easier and less error-prone. However the
format is very verbose and in its current form it does not seem
suitable for storing massive line lists such as individual lists with
more than 10$^{10}$ lines which are to be anticipated from the ExoMol
project. This will require the development of new protocols and,
presumably, adaptation of XSAMS.

Finally, we note that recent test of for models of water spectra in
hot Jupiter exoplanet \citep{jt523} suggest that pressure broadening
can have a significant influence, particularly at long wavelengths.
This means that pressure broadening parameters, particularly those
associated with collisions with H$_2$ should also be considered for
inclusion in the ExoMol database at some future date.

\section{Conclusion}

This paper lays out the scope and methodology for a new project,
ExoMol, whose aim is to provide comprehensive line lists of molecular
transition frequencies and probabilities. The major aim of this
project is to provide the necessary data to model atmospheres and
interpret spectra for exoplanets and cool stars. However it is
recognised that the line lists will have many other applications within
astrophysics and beyond. For example it is our practice not to exclude
transitions from our lists simply because they are too high in energy
to be thermally occupied.  This has already led to the identification
of new class of very vibrationally-hot water emissions in comets \citep{jt452}.
It is to be anticipated that such data will be important for assigning
and modelling maser emissions from high lying or hot states. Similarly
the database will be available for modelling what may be observable
in exoplanet characterisation missions such as the proposed Exoplanet
Characterisation Observatory (EChO) \citep{jt521} or FINESSE \citep{finesse}
space-borne telescopes.

The ExoMol project will generate very extensive line lists. These will be
documented in the present journal and deposited in the linked Strasbourg
data repository. The line lists, and other information about the project,
will also be made available via the ExoMol website, \url{www.exomol.com}.

\section*{Acknowledgements}

This work is supported by ERC Advanced Investigator Project 267219
and a Royal Society Wolfson Research Merit Award.
We thank those astronomers who have shared their insight with us in
the formulation of this project and in particular Giovanna Tinetti,
Peter Bernath, Caitlin Griffiths, Yakiv Pavlenko and Mark Swain. We
also thank the present members of the ExoMol team, Ala'a Azzam, Bob
Barber, Christian Hill, Lorenzo Lodi, Andrei Patrascu, Oleg Polyansky
and Clara Sousa-Silva, as well as Attila Cs\'{a}sz\'{a}r and Andrew
Kerridge for many helpful discussions on the project, and Olga Yurchenko
for her help.

\bibliographystyle{mn2e}

\label{lastpage}

\end{document}